\documentstyle[12pt]{article}
\begin{document}
\begin{titlepage}
\rightline{RU94-5-B}
\rightline{hep-th/9405406}
\vskip 4 truecm
\centerline{ON TESTING LOCAL QFT AT LHC \footnote{Talk delivered at  {\it Les
Rencontres de Physique de la Vallee d'Aoste:  Results and
Perspectives in Particle Physics.} La Thuile, Aosta, March 1994 }}
\vspace{24pt}
\baselineskip 12pt
\centerline{N.N. Khuri}
\vspace{0.4ex}
\centerline{\it Department of Physics}
\vspace{0.4ex}
\centerline{\it The Rockefeller University, New York, New York 10021}
\vspace{3mm}
\vspace{2.0in}
\begin{abstract}
We discuss the importance of measuring $\rho= ReF/ImF$ at LHC to
test the forward dispersion relations and local QFT.  It is pointed
out that at LHC we can reach a short distance domain that has not
been pre-explored by QED.  This is in contrast with all previous
tests of the dispersion relations.  We argue that the most likely
property of QFT to fail is polynomial boundedness.  In a theory
with `fundamental length', R, we study the consequences of having
exponential behavior in the amplitude of the form (exp$\pm i\sqrt{s}\;R)$,
as suggested by different models. We show that such a behavior makes a
significant and measurable contribution to $\rho$ even at energies
where $(\sqrt{s}\;R)$ is still small, O(0.1).
\end{abstract}
\end{titlepage}
\newpage
\baselineskip 18 pt

\hspace{.25in}For over three decades, the forward dispersion relations have
represented one of the few general rigorous consequences of local quantum field
theory.  Starting from the axiomatic formulation of QFT, one can
establish that the forward scattering amplitude, $F(s)$, has the following
properties: a.)  $F(s)$ is an analytic function of $s$ with two cuts
on the real
axis.  b.)  It satisfies the property of crossing symmetry.  c.)  The
optical theorem gives $ImF = k\sqrt{s} \sigma_{tot}$, where $k$ is
the center of mass momentum.  d.)  $F(s)$ is polynomially bounded
for large $|s|$ in the cut plane, $|F(s)|<C|s|^N$.  These four
properties and the Froissart-Martin bound lead to the dispersion
relations which enable us to calculate $ReF$ from the total
cross-section.$^{1)}$

Starting in 1960, these relations have been repeatedly tested at
practically every major new $pp$ or $\bar{p}p$ accelerator or
collider, covering a C.M. energy range of $\sqrt{s} = 7 GeV$ to
$\sqrt{s} = 550 GeV$.  As can be seen in figure 1, these tests
always led to a measured $(ReF/ImF)\equiv\rho$, which agreed with the
`theoretical' dispersion relations fit for $\rho$.  However,
all these tests were never preceded by a high level of expectation
that the results could have turned out to be different.  The reason
for that was the continuing improvement in the status of the
experimental and theoretical knowledge of QED.  For example when
$\rho$ was measured up to $\sqrt{s} = 7 GeV$, it was already
expected from the agreement between calculations and experiment
in QED, e.g. the $\mu$-meson magnetic moment, that there was no
breakdown in QFT at distances of order $(10\;GeV)^{-1}$.

At present we have the following situation.  If there is a
`fundamental length', $R$, then from QED and the results of the muon
magnetic moment we have

\begin{equation}
\alpha(m^2_\mu R^2)\leq 10^{-8};
\end{equation}
where the right hand side comes from the experimental error in the
muon moment.  This leads us to the following estimate

\begin{equation}
R^{-1}\geq O(100\;GeV).
\end{equation}
With model dependent arguments one can accommodate a fundamental
length $R$ such that $R^{-1}\approx 300-500\;GeV$, but not much better.
For our purposes here we can certainly make the following
conservative statement.  {\it Today, we have no experimental evidence
that can rule out the existence of a
fundamental length, $R$, such that $(R^{-1})>1\;TeV$}.$^{2)}$

Furthermore, we have already reached the end of the line, as far
as learning more from QED about $R$.  Magnetic moment calculations
to $O(\alpha^5)$ will not help since at that level hadronic and
electroweak contributions become significant.  Similarly, improving
our experimental error on the muon magnetic moment, will not help us
beyond $R^{-1}\approx O(1 TeV)$, since again we will have to deal with
contributions from non-electromagnetic processes that we cannot
accurately calculate.

With the construction of the $LHC$, $\sqrt{s}\cong 15\;TeV$, we will
be able to explore a short distance domain about which we have little
previous knowledge.  This is in sharp contrast to all the past tests
of the dispersion relations and QFT, which dealt with length scales
that had been pre-explored by QED.  In fact UA4/2 with
$\sqrt{s}\cong 550\;GeV$ was more or less on the dividing line.

There are four other factors which make the measurement of
$\rho$ at LHC even more compelling.  We list them here:

1.)  At $\sqrt{s}\approx 550\;GeV$, $\rho_{\bar{p}p}$ is small$^{3)}$,
$\rho_{\bar{p}p}= 0.135$.  In addition $\rho$ decreases
slowly (logarithmically).  The phase of $F$, $\phi$, with
$\phi_{\bar{p}p}=tan^{-1}(1/\rho)\approx 83^o$, at
$\sqrt{s}\approx 550\;GeV$.  Also for $\sqrt{s}>200\;GeV$ a good fit
for $\rho$ is $\rho\cong c[ln\;s]^{-1}$.  These facts will
turn out to be very useful below.

2.)  Cosmic ray data on $\sigma_{tot}$ indicate that $\sigma_{tot}(pp)$
continues to increase much beyond $s_{LHC}$.  It grows like $(ln\;s)^\alpha$.
While a range of $\alpha$`s are allowed, there is no doubt about the
fact that the increase continues.$^{4)}$ This limits the options we
have in the calculated value of $\rho$ at LHC energies.

3.)  At LHC we for the first time reach the region of large $\;ln\;s$.
Indeed we have $$ln(s_{LHC}/m^2_{proton}) \approx 17,$$ and powers
of $(ln\;s)$ differ significantly.  This puts us in a domain where the
phase relations$^{5)}$ derived by Kinoshita and the author become useful
tools.

4.)  From the theoretical point of view, there are recent results
which make the realm of superstring theories accessible to LHC.
Of importance here is the possible existence of a new internal
dimension at energies of a few TeV, i.e. low compared to the Planck
mass.  This new compact internal dimension could have a radius, $R$,
with $R^{-1} \approx O(1\;TeV)$, and space time given by $M_4\otimes S$.
The new dimension is supposed to lead to supersymmetry breakdown.
Such a suggestion, made earlier by several people, initially had several
problems.  However, it has recently been revived by Antoniadis$^{6)}$.  He
was able to present it in a form that not only overcomes most of the
difficulties but also converts it into a viable proposal.

The first question we must answer is how will the dispersion relations
fail. Which of the four properties of $F(s)$ listed in the first
paragraph of this paper will fail?  It is highly unlikely that
$F(s)$ in any viable theory will have complex singularities on the
physical sheet of the s-plane.  Indeed such singularities would
be a signal for unwanted physical states.  It is also highly unlikely
that crossing-symmetry fails.  The
third property of $F$, the optical theorem, is merely a statement on the
conservation of probability and thus on very solid ground.

The most likely point of failure is the property of polynomial
boundedness. This property is almost a direct consequence of the
axiom on tempered distributions.  It is the least physical of all
the axioms. We can defend our choice of polynomial boundedness as the most
likely pillar to fail by the following examples:

A. {\underline{String Theory}}

The work of Gross and Mende$^{7)}$ on scattering in string theory shows
that polynomial boundedness fails in that case but at energies
beyond the Planck scale.  A similar situation was discussed much
earlier when linearly rising Regge trajectories were first
introduced.$^{8)}$

It should be noted that in string theory analyticity, crossing, and the
optical theorem are not violated.

B. {\underline{Non-Local Potential Scattering}}

In the Schrodinger equation we replace the local interaction term by
a non-local one, i.e.,
\begin{equation}
V(|\vec{x}|)\psi(\vec{x})\rightarrow \int d^3y
V(|\vec{x}-\vec{y}|)\psi(\vec{y}).
\end{equation}
The non-local potential is chosen to vanish outside a sphere of radius $R$,
\begin{equation}
V(|\vec{x}-\vec{y}|)\equiv 0,\;\; {\rm{for}} \;\; |\vec{x}-\vec{y}|>R.
\end{equation}

Then one can still prove that the scattering amplitude, $F(s)$, will
still be analytic.  However,
polynomial boundedness is lost, and only
$|F(s).exp(ikR)|$ is polynomially bounded.$^{9)}$  Here
$k\equiv \sqrt{s}$.

We can define $F_{true}$ and $F_{false}$ as:
\begin{equation}
F_t(s)\equiv F_f(s)e^{-ikR}.
\end{equation}
$F_f(s)$ satisfies a dispersion relation, but Im${F_f} \not= k\sigma_{tot}$,
and hence not guaranteed to be positive.  However, for small $k$ such that
$(kR)<< 1$, $F_t(s)\cong F_f(s)$, and if we only do our experiments at
energies such that $\sqrt{s} << R^{-1}$, we will not detect any violation
of the dispersion relations or locality.

C. {\underline{Non-Local Field Theories}

Almost all non-local QFT proposals which appeared before the
introduction of the Wightman axioms had serious problems and violated
many established principles.  A noted exception is the more recent
random lattice method of T.D. Lee$^{10)}$ and collaborators, where
a fundamental length is introduced without creating difficulties.

Non-local field theories are expected to have exponential behavior
for $F(s)$ with a sign similar to that in the case of non-local
potential scattering if the non-locality occurs in the time component.

D. {\underline{New Compact Internal Dimension}}

In this case we have no violation of locality.  However, polynomial
boundedness could still breakdown due to the compact nature of the
extra dimension.  While the situation is not clear without a
concrete theory that can be analytically studied, one can still make
the following remark.  Namely, the sign in the exponential will
be the opposite to the one discussed above if polynomial boundedness
breaks down due to the extra dimension.  If we take seriously the
short distance behavior of the amplitude in x-space as given by
$\theta(x^2_o-\vec{x}^2-R^2\phi^2)$, where $\phi$ is the angle
on S, as compared to $\theta(x^2_o-\vec{x}^2)$ in the $M_4$
case, then we would have the situation where
\begin{equation}
F_t(s)=F_f(s)e^{i \frac{\sqrt{s}}{2} R}
\end{equation}
This will give a different and distinct signal for $\rho$ when
$\sqrt{s} R$ is non-negligible.

In a recent paper$^{11)}$ we have studied non-relativistic
potential scattering on $R_3\otimes S$, and we get some
significant changes in the properties of the forward scattering
amplitude that are due to the extra compact dimension.

In the rest of this paper we shall assume that polynomial
boundedness does indeed breakdown at some energy $R^{-1}$.  Taking
the specific ansatz, inspired by our non-local potentials, we
show numerically that even at energies such that
$(\sqrt{s} R)\approx (1/10)$ there will be a strong and observable
change in $\rho$.

We start with the ansatz:
\begin{equation}
F_t(s)\equiv F_f(s)e^{\mp i \frac{\sqrt{s}}{2} R}.
\end{equation}
Here $F_t$ is the `true' amplitude, and Im$F_t \equiv k\sqrt{s}\sigma_{tot}>0$.
On the other hand $F_f(s)$ is polynomially bounded, satisfies the
dispersion relation, but Im$F_f$ does not satisfy the positivity
condition.

Equation(7) looks at first sight like a tautology.  It is just a
definition of $F_f(s)$.  However, because of the special properties
of $\rho$ listed earlier, we can still learn some interesting
things which are experimentally quite relevant.  This will be
shown immediately.

There are two cases to be calculated separately.  They are
distinguished by the sign of the exponential in equation (7).

{\underline{Case I:}

Here we have,
\begin{equation}
F_t(s) = F_f(s) e^{-i \frac{\sqrt{s}}{2}R}.
\end{equation}
This is the case inspired by the example of non-local potentials.
There are two energy regions to consider, low energies, $\sqrt{s} R<<1$,
and a transitional region, $0.01 < \sqrt{s} R <0.4 < 1$.

A. {\underline{Low Energy Region:}

Here $(\sqrt{s} R) << 1$, and hence $F_t \cong F_f$ and $\rho_t\cong\rho_f$.

B.  {\underline{Transitional Region:}

This is defined as the energy range
\begin{equation}
0.01 < \sqrt{s} R \leq 0.4.
\end{equation}
Essentially it is the region where $\sqrt{s} R$ is non-negligible but
still below the threshold for producing excitations, $m^2_n = n^2/R^2$.
{}From eq. (8) we have,
\begin{equation}
ImF_f(s)=k\sqrt{s} \sigma_{tot}(s)[cos \frac{\sqrt{s}}{2}R+\rho_t(s)
sin \frac{\sqrt{s}}{2}R],
\end{equation}
which can be written as
\begin{equation}
Im F_f(s)=k\sqrt{s}\sigma_{tot}[1+ \rho_t(\frac{\sqrt{s}}{2}R)-\frac{1}{2}
{(\frac{\sqrt{s}}{2}R)}^2 + O({(\frac{\sqrt{s}}{2}R)}^3)].
\end{equation}

It is now clear that as long as $(\sqrt{s} R/2)< 0.2$, and $\rho_t <0.35$
(an assumption which we will show later is not needed), then in the
transitional
region we have
\begin{equation}
Im F_f=k\sqrt{s} \sigma_{tot}[1-O(2\%)],
\end{equation}
with $\sqrt{s} R<0.4$.

Since $F_f(s)$ satisfies the dispersion relation, we can now extend the
results into the transitional region where the optical theorem is still
approximately good, and get
\begin{equation}
\rho_f(s)\cong \rho_{fit}(s), \;\;\; {\rm{for}} \;\;\; 0.01<\sqrt{s} R\leq0.3.
\end{equation}
Here $\rho_{fit}$ is the value of $\rho$ obtain from the standard dispersion
relation fit shown in figure 1.

The true and false phases are related by
\begin{equation}
\phi_t(s)=\phi_f(s)-\frac{\sqrt{s}}{2} R.
\end{equation}
But $\phi_f\equiv tan^{-1}(1/\rho_f)$, and in the transitional
region using (13) we get
\begin{equation}
\phi_t(s)=tan^{-1}(1/\rho_{fit})-\frac{\sqrt{s}}{2} R;\;\;\; 0.01<\sqrt{s}
R<0.3.\end{equation}

Using $\rho_t\equiv cot\phi_t$, we get $\rho_t$ in the transitional
region.  The result is shown in figure 1, for the case where we
choose $(R^{-1})= 12\;TeV$.  Even with that small length, by the time
$\sqrt{s} =4\;TeV$ the value of $\rho_t$ is about 1/3, almost 2.5 times
larger than UA4/2.  The remarkable thing is that this occurs even
when $\sqrt{s} R\leq 1/3$.  In fact we begin to see a measurable
effect on $\rho$ when $\sqrt{s} R\approx 0.1$.  Hence one gets a signal
even when $\sqrt{s}$ is an order of magnitude below $(R^{-1})$.

Finally, we show that the assumption about $\rho_t$ which we made in
estimating the r.h.s. of eq. (11) and arriving at eq. (12) is not
needed.  All one need to do is to divide the interval
$0.01R^{-1}\leq\sqrt{s}\leq 0.4R^{-1}$ into ten intervals and
carry out the calculations from Eqs. (11)-(15) repeatedly starting from
$\sqrt{s} =0.01R^{-1}$.  This way we can guarantee that the estimate
(12) is correct throughout the transitional region.

{\underline{Case II:}

This is the situation that will obtain if polynomial boundedness
is broken by an extra internal dimension.  Here we have
\begin{equation}
F_t(s)=F_f(s)\;e^{+i\frac{\sqrt{s}}{2} R}.
\end{equation}

Again $\rho_f\cong \rho_{fit}$ for $0.01<\sqrt{s} R<0.3$, but now
$\rho_t$ decreases and indeed becomes negative. The result is shown in fig.1.
Again the signal is remarkable and the fact that $\rho_t$ becomes
negative allows us to make use of the inequalities of ref. 5.

In conclusion we state that the exponential behavior of $F(s)$
leads to a phase that is linearly dependent on $\sqrt{s}$,
unlike the case where polynomial boundedness leads to a
logarithmic behavior of $argF$.  Regardless, of the result a
measurement of $\rho_{pp}$ at LHC energies will be quite
important.  Either $\rho_{expt} = \rho_{fit}$ at LHC, which
effectively gives us a very high precision test of locality
in QFT, $R^{-1} = O(10^2)\;TeV$ or $R=O(10^{-19}) cms$.  A
result QED cannot match.  Or, we get a disagreement between
$\rho_{expt}$ and the dispersion relation result which would
mean a breakdown in QFT, and possibly suggestions of new
structures in space-time at short distances.
\section*{Acknowledgements}
The author thanks M. Evans and H.-C. Ren for several helpful remarks.
This work was supported in part by the U.S. Department of
Energy under grant no. DOE91ER40651 TaskB.


\begin{thebibliography}{99}
\baselineskip 12 pt

\bibitem{}
The first proof of the dispersion relations is due to N.N. Bogoliubov,
see N.N. Bogoliubov and D.V. Shirkov, {\it Introduction to
the Theory of Quantized Fields}, (Interscience Publishers, London,
1959).
\vspace {-.1 truecm}

\bibitem{}
T. Kinoshita, {\it Quantum Electrodynamics}, T.K. Editor,
(World Scientific, 1990), pp. 471.
\vspace {-.1 truecm}

\bibitem{}
{Giorgio Matthiae, these proceedings (l993 edition), pp. 177-187.}
\vspace {-.1 truecm}

\bibitem {Kowitt}
T.K. Gaisser, U.P. Sukhatme and G.B. Yodh, Phys. Rev. {\bf D36},
1350 (l987).
\vspace {-.1 truecm}

\bibitem{}
N.N. Khuri and T. Kinoshita, Phys. Rev. {\bf B140}, 706 (l965).
\vspace {-.1 truecm}

\bibitem{}
I. Antoniadis, C. Munoz and M. Quiros, Nucl Phys. {\bf B397}, 515
(l993).
\vspace {-.1 truecm}

\bibitem{}
D.J. Gross and P. Mende, Nucl. Phys. {\bf B303}, 407 (l988).
\vspace {-.1 truecm}

\bibitem{}
N.N. Khuri, Phys. Rev. Lett. {\bf 18}, 1094 (l967).
\vspace {-.1 truecm}

\bibitem{}
N.N. Khuri, Thesis (Princeton University, 1957), unpublished.
\vspace {-.1 truecm}

\bibitem{}
R. Friedberg and T.D. Lee, Nucl. Phys. {\bf D225}, 1 (l983).
\vspace {-.1 truecm}

\bibitem{}
N.N. Khuri, ``Potential Scattering on $R_3\otimes S$'', preprint
in preparation.
\vspace {-.1 truecm}

\end{thebibliography}
\end{document}